# Neutrino Mixing and Leptonic *CP* Phase in Neutrino Oscillations

D. A. Ryzhikh* and K. A. Ter-Martirosyan**

*Institute of Theoretical and Experimental Physics, ul. Bol'shaya Cheremushkinskaya 25, Moscow, 117259 Russia*
*\* e-mail: ryzhikh@heron.itep.ru*
*\*\* e-mail: termarti@heron.itep.ru*
Received December 15, 2000; in final form, July 2, 2001

Oscillations of the Dirac neutrinos of three generations in vacuum are considered with allowance made for the effect of the *CP*-violating leptonic phase (analogue of the quark *CP* phase) in the lepton mixing matrix. The general formulas for the probabilities of neutrino transition from one sort to another in oscillations are obtained as functions of three mixing angles and the *CP* phase. It is found that the leptonic *CP* phase can, in principle, be reconstructed by measuring the oscillation-averaged probabilities of neutrino transition from one sort to another. The manifestation of the *CP* phase as a deviation of the probabilities of direct processes from those of inverse processes is an effect that is practically unobservable as yet. © *2001 MAIK "Nauka/Interperiodica"*.

Rapidly growing interest in neutrino physics has recently been stimulated by new data from facilities at Kamiokande [1], Super-Kamiokande [2], LSND [3], CHOOZ [4], and some others [5, 6]. These data are indicative, directly or indirectly, of the Pontecorvo (vacuum) oscillations [7] of neutrinos of three types, $\nu_e$, $\nu_\mu$, and $\nu_\tau$. The frequency of these oscillations, i.e., of the $\nu_i \rightleftarrows \nu_k$ transitions in oscillations, is proportional to $\sin^2\left(\frac{m_3^2 - m_2^2}{4p_\nu}t\right)$, where $p_\nu$ is the ultrarelativistic neutrino momentum and $t = L/c$ is the time it takes for a neutrino to run from source to detector (the distance $L$ is called the "base length"). In what follows, we set $c = 1$ for convenience.

The presence of vacuum oscillations means that

(i) similar to quarks, neutrinos produced in decays or collisions have no definite mass [7] but are the superpositions of neutrinos $\nu_1^\circ$, $\nu_2^\circ$, and $\nu_3^\circ$, which have small ($m_i \sim 10^{-2}$–$10^{-4}$ eV) through definite masses [8, 9]:

$$\nu_\alpha(\mathbf{x}, t) = \sum_{i=1}^{3} \hat{V}_{\alpha i}^l \nu_i^\circ(\mathbf{x}, t), \quad (1)$$
$$\alpha = e, \mu, \tau; \quad i = 1, 2, 3.$$

(ii) neutrinos of different generations have different masses; i.e., $m_i^2 - m_k^2 \neq 0$.

It is assumed in Eq. (1) that neutrinos $\nu_i^\circ$ move with ultrarelativistic energy $E_i = \sqrt{\mathbf{p}_\nu^2 + m_i^2} \simeq |\mathbf{p}_\nu| + m_i^2/2p_\nu$ in a beam along the $X$ axis of their momentum. Therefore, at the detection time $t_0 = L$ we have

$$\nu_i(\mathbf{x}, t) = \nu_i(L, t_0) = \exp(i\mathbf{p}_\nu \mathbf{x})\exp(-iE_i t)\nu_i^\circ(0)$$
$$\simeq \exp\left(-i\frac{m_i^2}{2p_\nu}t\right)\nu_i^\circ(0) = \exp(-i\varphi_i t)\nu_i^\circ(0), \quad (2)$$
$$\varphi_i = \frac{m_i^2}{2p_\nu}.$$

The Maki–Nagava–Sakata (MNS) mixing matrix $\hat{V}_{\alpha i}^l$ of Dirac neutrinos [9] has the same form as the SKM mixing matrix of quarks [10], but with its own mixing angles $\vartheta_{12}$, $\vartheta_{13}$, and $\vartheta_{23}$ and its own *CP*-violating phase $\delta_l$:

$$\hat{V}^l$$

$$= \begin{pmatrix} c_{12}c_{13} & s_{12}c_{13} & s_{13}e^{-i\delta_l} \\ -s_{12}c_{13} - c_{12}s_{23}s_{13}e^{i\delta_l} & c_{12}c_{23} - s_{12}s_{23}s_{13}e^{i\delta_l} & s_{23}c_{13} \\ s_{12}s_{23} - c_{12}c_{23}s_{13}e^{i\delta_l} & -c_{12}s_{23} - s_{12}c_{23}s_{13}e^{i\delta_l} & c_{23}c_{13} \end{pmatrix}, \quad (3)$$

where $s_{ik} = \sin\vartheta_{ik}$ and $c_{ik} = \cos\vartheta_{ik}$. Like $\hat{V}_{CKM}$, the matrix $\hat{V}^l$ is unitary; i.e., $\hat{V}^l \hat{V}^{l+} = 1$.

Previous analysis of experimental data [1–6] gave the following mixing angles [8, 9, 11]

$$\vartheta_{12} = (42.1 \pm 6.9)°, \quad \vartheta_{13} = (2.3 \pm 0.6)°,$$
$$\vartheta_{23} = (43.6 \pm 3.1)° \quad (4)$$

for $\delta_l = 0$ and

$$m_3 = (0.058 \pm 0.025)\ \text{eV},$$
$$m_2 = (0.0060 \pm 0.0035)\ \text{eV}, \quad m_1 \ll m_2. \quad (5)$$

Here, the mean error in the mixing angles and the masses is taken from figures and tables in [8, 9, 11], where the CHOOZ data [4] from ground-based $\nu_e$ sources were taken into account.[1]

Below, we will consider the possibility of determining the leptonic CP phase $\delta_l$ from the data obtained in the today and immediate-future experiments of the type [1–6], where oscillations were not observed directly but only the oscillation-averaged probabilities $P(\nu_\alpha \nu_\beta)$ of $\nu_\alpha \longrightarrow \nu_\beta$ transitions were measured.

The action of matrix (3) on the column vector $\nu_i^\circ$ gives [see Eq. (1)]

$$\begin{pmatrix} \nu_e \\ \nu_\mu \\ \nu_\tau \end{pmatrix}_t = \hat{V}_l \begin{pmatrix} \nu_1^\circ \\ \nu_2^\circ \\ \nu_3^\circ \end{pmatrix},$$

i.e.,

$$\nu_e(t) = [c_{12}c_{13}\nu_1^\circ(0) + s_{12}c_{13}\nu_2^\circ(0)e^{-i\varphi_{21}}$$
$$+ s_{13}\nu_3^\circ(0)e^{-i\varphi_{31} - i\delta_l}]\exp\left(-i\frac{m_1^2}{2p_\nu}t\right),$$

$$\nu_\mu(t) = [-(s_{12}c_{13} + c_{12}s_{23}s_{13}e^{i\delta_l})\nu_1^\circ(0)$$
$$+ (c_{12}c_{23} - s_{12}s_{23}s_{13}e^{i\delta_l})\nu_2^\circ(0)e^{-i\varphi_{21}} \quad (6)$$
$$+ c_{13}s_{23}\nu_3^\circ(0)e^{-i\varphi_{31}}]\exp\left(-i\frac{m_1^2}{2p_\nu}t\right),$$

$$\nu_\tau(t) = [(s_{12}s_{23} - c_{12}c_{23}s_{13}e^{i\delta_l})\nu_1^\circ(0)$$
$$- (c_{12}s_{23} + s_{12}c_{23}s_{13}e^{i\delta_l})\nu_2^\circ(0)e^{-i\varphi_{21}}$$
$$+ c_{13}c_{23}\nu_3^\circ(0)e^{-i\varphi_{31}}]\exp\left(-i\frac{m_1^2}{2p_\nu}t\right),$$

where, taking into account the dependence (2) of neutrino states on time $t = L$, one has

$$\varphi_{ij} = \frac{(m_i^2 - m_j^2)}{2p_\nu}t = 2.54 \frac{(m_i^2 - m_j^2)(\text{eV}^2)}{2E_\nu(\text{MeV})} L(\text{m}), \quad (7)$$

with $E_\nu \simeq p_\nu$ being the neutrino energy in a beam, $E_\nu \gg m_3 > m_2 > m_1$. Because the neutrino states $\nu_i^\circ(0)$ are orthonormalized, i.e., $(\overline{\nu_i^\circ}\nu_k^\circ) = \delta_{ik}$,[2] one has for the amplitudes $A_{a\to b} = (\overline{\nu_\beta(t)}\nu_\alpha(0))$ and probabilities $P(\nu_\alpha\nu_\beta) = |(\overline{\nu_\beta(t)}\nu_\alpha(0))|^2$ of $\nu_\alpha(0) \longrightarrow \nu_\beta(t)$ transitions in vacuum

$$P(\nu_e\nu_e) = |c_{12}^2c_{13}^2 + s_{12}^2c_{13}^2 e^{i\varphi_{21}} + s_{13}^2 e^{i\varphi_{31}}|^2,$$

$$P(\nu_\mu\nu_\mu) = ||c_{13}s_{12} + c_{12}s_{13}s_{23}e^{i\delta_l}|^2$$
$$+ |c_{12}c_{23} - s_{12}s_{23}s_{13}e^{i\delta_l}|^2 e^{i\varphi_{21}} + c_{13}^2s_{23}^2 e^{i\varphi_{31}}|^2, \quad (8)$$

$$P(\nu_\tau\nu_\tau) = ||s_{12}s_{23} - c_{12}c_{23}s_{13}e^{i\delta_l}|^2$$
$$+ |c_{12}s_{23} + s_{12}c_{23}s_{13}e^{i\delta_l}|^2 e^{i\varphi_{21}} + c_{13}^2c_{23}^2 e^{i\varphi_{31}}|^2,$$

and

$$P(\nu_e\nu_\mu) = |c_{12}c_{13}(c_{13}s_{12} + c_{12}s_{23}s_{13}e^{i\delta_l})$$
$$- c_{13}s_{12}(c_{12}c_{23} - s_{12}s_{23}s_{13}e^{i\delta_l})e^{i\varphi_{21}}$$
$$- s_{13}c_{13}s_{23}e^{i(\delta_l + \varphi_{31})}|^2,$$

$$P(\nu_e\nu_\tau) = |c_{12}c_{13}(s_{23}s_{12} - c_{12}c_{23}s_{13}e^{i\delta_l})$$
$$- c_{13}s_{12}(c_{12}c_{23} + c_{23}s_{12}s_{13}e^{i\delta_l})e^{i\varphi_{21}} \quad (9)$$
$$+ s_{13}c_{13}c_{23}e^{i(\delta_l + \varphi_{31})}|^2,$$

$$P(\nu_\mu\nu_\tau) = |(c_{13}s_{12} + c_{12}s_{13}s_{23}e^{i\delta_l})$$
$$\times (s_{12}s_{23} - c_{12}c_{23}s_{13}e^{-i\delta_l})$$
$$+ (c_{12}c_{23} - s_{12}s_{13}s_{23}e^{i\delta})$$
$$\times (c_{12}s_{23} + c_{23}s_{12}s_{13}e^{i\delta})e^{i\varphi_{21}} - c_{23}^2c_{13}^2s_{23}^2 e^{i\varphi_{31}}|^2.$$

---

[1] Unfortunately, results (4) and (5) obtained in [8, 9, 11] using the data reported in [1–6] are insufficiently reliable, especially those based on the data for solar and, partly, atmospheric electron neutrinos $\nu_e$, whose interaction with matter within the Sun or Earth can invert their spin and transform $(\nu_e)_L$ to the sterile, i.e., noninteracting, state $(\nu_e)_R$. This so-called MSW effect [12] does not occur in an analysis of the data from ground-based $\nu_e$ sources, e.g., CHOOZ data [4], whose processing yields very small angles $\vartheta_{13} \sim 2°$–$3°$ [see Eq. (4)].

[2] For the Majorana neutrinos, whose fields $\nu_i^0(0)$ with definite mass are real, the requirement that fields $\nu_e$, $\nu_\mu$, and $\nu_\tau$ (6) produced in the weak interaction be real even at $t = 0$ would mean that real matrix (3) is orthogonal, i.e., $\delta_l = \delta_{13} = \pi$ or 0. This is also true for the phases $\delta_{12}$ and $\delta_{23}$, which are omitted in Eqs. (3) and (6) because they lead to vanishingly small probabilities of the $\nu_i \longrightarrow \bar{\nu}_k$ transitions with amplitudes $\sim m_\nu/E \sim 10^{-6}$–$10^{-9}$. However, besides simplicity and aesthetics, there are no other reasons for requiring that fields (6) be real and CP phase be zero. We intend to consider the Majorana neutrino oscillations elsewhere [13].

Averaging these probabilities over oscillations, i.e., over phases $\varphi_{ij}$ [by setting $\langle \sin^2\varphi_{ij}\rangle = \langle \cos^2\varphi_{ij}\rangle = 1/2$ and $\langle \cos(\varphi_{ij} \pm \delta_l)\rangle = \langle \cos\varphi_{ij}\rangle = 0$ in Eqs. (8) and (9)], we obtain the following energy-independent probabilities, which were only measured to date [1–6]:

$$\langle 1 - P(\nu_e\nu_e)\rangle = A_{ee},$$
$$\langle 1 - P(\nu_\mu\nu_\mu)\rangle = A_{\mu\mu} + B_{\mu\mu}\cos\delta_l + C_{\mu\mu}\cos^2\delta_l,$$
$$\langle 1 - P(\nu_\tau\nu_\tau)\rangle = A_{\tau\tau} + B_{\tau\tau}\cos\delta_l + C_{\tau\tau}\cos^2\delta_l,$$
$$\langle P(\nu_e\nu_\mu)\rangle = A_{e\mu} + B_{e\mu}\cos\delta_l,$$
$$\langle P(\nu_e\nu_\tau)\rangle = A_{e\tau} + B_{e\tau}\cos\delta_l,$$
$$\langle P(\nu_\mu\nu_\tau)\rangle = A_{\mu\tau} + B_{\mu\tau}\cos\delta_l + C_{\mu\tau}\cos(2\delta_l),$$
(10)

where

$$A_{ee} = \frac{1}{2}[c_{13}^4\sin^2(2\vartheta_{12}) + \sin^2(2\vartheta_{13})],$$

$$A_{\mu\mu} = \frac{1}{2}[(c_{13}^2 + (c_{12}^4 + s_{12}^4)s_{13}^2)\sin^2(2\vartheta_{23})$$
$$+ (s_{13}^4\sin^2(2\vartheta_{12}) + \sin^2(2\vartheta_{13}))s_{23}^4 + c_{23}^4\sin^2(2\vartheta_{12})],$$

$$B_{\mu\mu} = \frac{1}{2}(c_{23}^2 - s_{23}^2 s_{13}^2) s_{13} \sin(2\vartheta_{23})\sin(4\vartheta_{12}),$$

$$C_{\mu\mu} = -\frac{1}{2}s_{13}^2\sin^2(2\vartheta_{23})\sin^2(2\vartheta_{12});$$

$$A_{\tau\tau} = \frac{1}{2}[(c_{13}^2 + (c_{12}^4 + s_{12}^4)s_{13}^2)\sin^2(2\vartheta_{23})$$
$$+ (s_{13}^4\sin^2(2\vartheta_{12}) + \sin^2(2\vartheta_{13}))c_{23}^4 + s_{23}^4\sin^2(2\vartheta_{12})],$$

$$B_{\tau\tau} = -\frac{1}{2}s_{13}\sin(2\vartheta_{23})(s_{23}^2 - c_{23}^2 s_{13}^2)\sin(4\vartheta_{12}),$$

$$C_{\tau\tau} = -\frac{1}{2}s_{13}^2\sin^2(2\vartheta_{23})\sin^2(2\vartheta_{12});$$
(11)

$$A_{e\mu} = \frac{1}{4}[(1 + c_{12}^4 + s_{12}^4)s_{23}^2\sin^2(2\vartheta_{13})$$
$$+ 2c_{13}^2 c_{23}^2\sin^2(2\vartheta_{12})],$$

$$B_{e\mu} = \frac{1}{8}c_{13}\sin(2\vartheta_{13})\sin(2\vartheta_{23})\sin(4\vartheta_{12});$$

$$A_{e\tau} = \frac{1}{4}[(1 + c_{12}^4 + s_{12}^4)c_{23}^2\sin^2(2\vartheta_{13})$$
$$+ 2c_{13}^2 s_{23}^2\sin^2(2\vartheta_{12})],$$

$$B_{e\tau} = -\frac{1}{8}c_{13}\sin(2\vartheta_{13})\sin(2\vartheta_{23})\sin(4\vartheta_{12});$$

$$A_{\mu\tau} = \frac{1}{4}[2s_{13}^2\sin^2(2\vartheta_{12})\cos^2(2\vartheta_{23})$$
$$+ \sin^2(2\vartheta_{23})\{(c_{12}^4 + s_{12}^4)s_{13}^4 + c_{13}^4 + c_{12}^4 + s_{12}^4\}],$$

$$B_{\mu\tau} = \frac{1}{8}(1 + s_{13}^2)s_{13}\sin(4\vartheta_{12})\sin(4\vartheta_{23}),$$

$$C_{\mu\tau} = -\frac{1}{4}s_{13}^2\sin^2(2\vartheta_{12})\sin^2(2\vartheta_{23}).$$

Note that the obvious relationships

$$1 - P(\nu_\alpha\nu_\alpha) = P(\nu_\alpha\nu_\beta) + P(\nu_\alpha\nu_\gamma), \quad \alpha, \beta, \gamma = e, \mu, \tau,$$

are satisfied, and $P(\nu_\beta\nu_\alpha) = [P(\nu_\alpha\nu_\beta)]_{\delta_l \to -\delta_l}$. Using the general formulas for oscillation probabilities from the Appendix, we obtain the following expressions for the differences between the probabilities of forth-back neutrino transitions:

$$P(\nu_\mu\nu_e) - P(\nu_e\nu_\mu)$$
$$= a_0(\sin\varphi_{21} + \sin\varphi_{32} - \sin\varphi_{31})\sin\delta_l,$$

$$P(\nu_\tau\nu_e) - P(\nu_e\nu_\tau)$$
$$= -a_0(\sin\varphi_{21} + \sin\varphi_{32} - \sin\varphi_{31})\sin\delta_l,$$
(12)

$$P(\nu_\tau\nu_\mu) - P(\nu_\mu\nu_\tau)$$
$$= a_0\left(\sin\varphi_{21} - 2\sin\frac{\varphi_{21}}{2}\cos\frac{(\varphi_{31} + \varphi_{32})}{2}\right)\sin\delta_l,$$

where $a_0 = \frac{1}{2}c_{13}\sin 2\vartheta_{12}\sin 2\vartheta_{13}\sin 2\vartheta_{23} \simeq 0.07$. Unfortunately, the phases $\varphi_{ik}$ appearing in these relationships depend on the neutrino energy in a beam; therefore, to determine the $\sin\delta_l$ value from Eq. (12), neutrinos $\nu_\alpha$ and $\nu_\beta$ should have the same energy $E$ in an experiment. Modern beams include only continuous-spectrum neutrinos, and the effect reflected in Eq. (12) vanishes after averaging over the phases $\varphi_{ik}$.

However, the *CP* phase can be obtained in a different way by using Eqs. (10) and (11) and the experimental data similar to those obtained in [1–6] but having higher accuracy in order to compensate the smallness of angle $\vartheta_{13}$. For clearness, let us introduce the coefficients $b_{ik} = B_{ik}/A_{ik}$ and $c_{ik} = C_{ik}/A_{ik}$ in Eqs. (10) and (11). Because of the smallness of $s_{13} = \sin\vartheta_{13} \simeq 0.07$, almost all of these coefficients are very small and are on the order of a fraction of percent:

$$A_{ee} = 0.499;$$
$$A_{\mu\mu} = 0.636, \quad b_{\mu\mu} = 0.0058, \quad c_{\mu\mu} = -0.0038;$$
$$A_{\tau\tau} = 0.613, \quad b_{\tau\tau} = 0.0055, \quad c_{\tau\tau} = -0.0040;$$
$$A_{e\mu} = 0.261, \quad b_{e\mu} = 0.014;$$
(13)

$$A_{e\tau} = 0.238, \quad b_{e\tau} = -0.015;$$

$$A_{\mu\tau} = 0.373, \quad b_{\mu\tau} = 0.0005, \quad c_{\mu\tau} = -0.0032.$$

For this reason, the ratio of the number of produced $\nu_\mu$ to that of $\nu_\tau$ in the primary $\nu_e$ beam at large distances $L$ (about 300–500 km) will weakly decrease with increasing $\delta_l$ from 0 to $\pi$:

$$\frac{N_{\nu_\mu}}{N_{\nu_\tau}} = \frac{\langle P(\nu_e \nu_\mu)\rangle}{\langle P(\nu_e \nu_\tau)\rangle} = \simeq \frac{A_{e\mu}}{A_{e\tau}}(1 + (b_{e\mu} - b_{e\tau})\cos\delta_l), \quad (14)$$

where $b_{e\mu} - b_{e\tau} \simeq 2b_{e\mu} \simeq 2.8\%$ and $A_{e\mu}/A_{e\tau} \simeq 1.04$. Therefore, if the experimentally measured ratio (14) differs from $1.04 + 0.03 = 1.07$ by more than 1–3%, this would indicate that $\cos\delta_l < 1$; i.e., $\delta_l \neq 0$.

If $s_{13} = \sin\vartheta_{13} > 0.07$, i.e., if $s_{13}$ is larger than the value used in this work, then the *CP* phase will be manifested more strongly. In particular, for $\vartheta_{13} = 14°$, we have $N_{\nu_\mu}/N_{\nu_\tau} \simeq 1.07(1 + 0.08\cos\delta_l)$, and the coefficient $a_0$ in Eq. (12) is $a_0 \simeq 0.23$.

We are grateful to D.I. Kazakov for information about last-year Osaka Conference and S.P. Mikheev for discussion of the current situation in neutrino physics. This study was supported by the Russian Foundation for Basic Research (project nos. 00-15-96786 and 00-02-16363).

*APPENDIX*

The probabilities of all neutrino transitions in vacuum in the Pontecorvo oscillations with allowance made for the *CP* phase $\delta_l$ are determined by the following general algebraic formulas:

$$1 - P(\nu_e \nu_e) = c_{12}^2 \sin^2(2\vartheta_{13})\sin^2(\varphi_{31}/2) \quad (A1)$$
$$+ c_{13}^4 \sin^2(2\vartheta_{12})\sin^2(\varphi_{21}/2) + s_{12}^2 \sin^2(2\vartheta_{13})\sin^2(\varphi_{32}/2),$$

$$1 - P(\nu_\mu \nu_\mu) = \{c_{23}^4 \sin^2(2\vartheta_{12}) + s_{12}^4 s_{13}^2 \sin^2(2\vartheta_{23})$$
$$+ s_{23}^4 s_{13}^4 \sin^2(2\vartheta_{12}) + c_{12}^4 s_{13}^2 \sin^2(2\vartheta_{23})$$
$$+ \cos\delta_l \sin(4\vartheta_{12})\sin(2\vartheta_{23})(s_{13} c_{23}^2 - s_{13}^3 s_{23}^2)$$
$$- \cos^2\delta_l s_{13}^2 \sin^2(2\vartheta_{23})\sin^2(2\vartheta_{12})\} \sin^2(\varphi_{21}/2) \quad (A2)$$
$$+ \{s_{12}^2 c_{13}^2 \sin^2(2\vartheta_{23}) + c_{12}^2 s_{23}^4 \sin^2(2\vartheta_{13}) + \cos\delta_l s_{23}^2 c_{13}$$
$$\times \sin(2\vartheta_{12})\sin(2\vartheta_{23})\sin(2\vartheta_{13})\} \sin^2(\varphi_{31}/2)$$
$$+ \{c_{12}^2 c_{13}^2 \sin^2(2\vartheta_{23}) + s_{12}^2 s_{23}^4 \sin^2(2\vartheta_{13}) - \cos\delta_l s_{23}^2 c_{13}$$
$$\times \sin(2\vartheta_{12})\sin(2\vartheta_{23})\sin(2\vartheta_{13})\} \sin^2(\varphi_{32}/2),$$

$$1 - P(\nu_\tau \nu_\tau) = \{c_{23}^4 \sin^2(2\vartheta_{12}) + s_{12}^4 s_{13}^2 \sin^2(2\vartheta_{23})$$
$$+ c_{23}^4 s_{13}^4 \sin^2(2\vartheta_{12}) + c_{12}^4 s_{13}^2 \sin^2(2\vartheta_{23})$$
$$+ \cos\delta_l \sin(4\vartheta_{12})\sin(2\vartheta_{23})(s_{13} c_{23}^2 - s_{13}^3 s_{23}^2)$$
$$- \cos^2\delta_l s_{13}^2 \sin^2(2\vartheta_{23})\sin^2(2\vartheta_{12})\} \sin^2(\varphi_{21}/2) \quad (A3)$$
$$+ \{s_{12}^2 c_{13}^2 \sin^2(2\vartheta_{23}) + c_{12}^2 s_{23}^4 \sin^2(2\vartheta_{13}) + \cos\delta_l c_{23}^2 c_{13}$$
$$\times \sin(2\vartheta_{12})\sin(2\vartheta_{23})\sin(2\vartheta_{13})\} \sin^2(\varphi_{31}/2)$$
$$+ \{c_{12}^2 c_{13}^2 \sin^2(2\vartheta_{23}) + s_{12}^2 c_{23}^4 \sin^2(2\vartheta_{13}) + \cos\delta_l c_{23}^2 c_{13}$$
$$\times \sin(2\vartheta_{12})\sin(2\vartheta_{23})\sin(2\vartheta_{13})\} \sin^2(\varphi_{32}/2),$$

$$P(\nu_e \nu_\mu) = \frac{1}{4}\{\sin^2(2\vartheta_{13})(s_{23}^2 + c_{12}^4 s_{23}^2 + s_{12}^4 s_{23}^2)$$
$$+ \frac{1}{2}c_{13}\sin(2\vartheta_{13})\sin(2\vartheta_{23})\sin(4\vartheta_{12})\cos\delta_l$$
$$- 2c_{13}^2 \sin^2(2\vartheta_{12})(c_{23}^2 - s_{13}^2 s_{23}^2)\cos(\varphi_{21})$$
$$- 2s_{23}^2 \sin^2(2\vartheta_{13})(c_{12}^2 \cos(\varphi_{31}) + s_{12}^2 \cos(\varphi_{32})) \quad (A4)$$
$$+ c_{13}\sin(2\vartheta_{12})\sin(2\vartheta_{13})\sin(2\vartheta_{23})(s_{12}^2 \cos(\delta_l + \varphi_{21})$$
$$- c_{12}^2 \cos(\delta_l - \varphi_{21})) + c_{13}\sin(2\vartheta_{12})\sin(2\vartheta_{13})\sin(2\vartheta_{23})$$
$$\times (\cos(\delta_l + \varphi_{32}) - \cos(\delta_l - \varphi_{31})) + 2c_{13}^2 c_{23}^2 \sin^2(2\vartheta_{12})\},$$

$$P(\nu_e \nu_\tau) = \frac{1}{4}\{\sin^2(2\vartheta_{13})(c_{23}^2 + c_{12}^4 c_{23}^2 + s_{12}^4 c_{23}^2)$$
$$- \frac{1}{2}c_{13}\sin(2\vartheta_{13})\sin(2\vartheta_{23})\sin(4\vartheta_{12})\cos\delta_l$$
$$+ 2c_{13}^2 \sin^2(2\vartheta_{12})(c_{23}^2 - s_{13}^2 s_{23}^2)\cos(\varphi_{21})$$
$$- 2c_{23}^2 \sin^2(2\vartheta_{13})(c_{12}^2 \cos(\varphi_{31}) + s_{12}^2 \cos(\varphi_{32})) \quad (A5)$$
$$+ c_{13}\sin(2\vartheta_{12})\sin(2\vartheta_{13})\sin(2\vartheta_{23})(c_{12}^2 \cos(\delta_l - \varphi_{21})$$
$$- s_{12}^2 \cos(\delta_l + \varphi_{21})) + c_{13}\sin(2\vartheta_{12})\sin(2\vartheta_{13})\sin(2\vartheta_{23})$$
$$\times (\cos(\delta_l + \varphi_{31}) - \cos(\delta_l + \varphi_{32})) + 2c_{13}^2 s_{23}^2 \sin^2(2\vartheta_{12})\},$$

$$P(\nu_\mu \nu_\tau) = \frac{1}{4}\{2s_{13}^2 \sin^2(2\vartheta_{12})\cos^2(2\vartheta_{23})$$
$$+ (c_{13}^4 + c_{12}^4 + s_{12}^4 + (c_{12}^4 + s_{12}^{4r}) s_{13}^4)\sin^2(2\vartheta_{23})$$
$$- [2s_{13}^2 (c_{23}^4 + s_{23}^4) \sin^2(2\vartheta_{12}) + [2s_{13}^2 (c_{12}^4 + s_{12}^4)$$
$$- (1 + s_{13}^4)\sin^2(2\vartheta_{12})]\sin^2(2\vartheta_{23})]\cos(\varphi_{21})$$
$$- \Big[2c_{13}^2(s_{12}^2 + c_{12}^2 s_{13}^2)\sin^2(2\vartheta_{23})$$
$$- \frac{1}{2}c_{13}\sin(2\vartheta_{12})\sin(2\vartheta_{13})\sin(4\vartheta_{23})\cos\delta_l\Big]\cos(\varphi_{31})$$

$$+ \left[ 2c_{13}^2 \sin^2(2\vartheta_{23})(s_{12}^2 s_{13}^2 - c_{12}^2) \right. \quad (A6)$$

$$\left. - \frac{1}{2} c_{13} \sin(2\vartheta_{12}) \sin(2\vartheta_{13}) \sin(4\vartheta_{23}) \cos\delta_l \right] \cos(\varphi_{32})$$

$$+ 2c_{13} \sin(2\vartheta_{12}) \sin(2\vartheta_{13}) \sin(2\vartheta_{23})$$

$$\times \sin\delta_l \sin(\varphi_{21}/2) \cos\left(\frac{\varphi_{31} + \varphi_{32}}{2}\right)$$

$$+ s_{13} \sin(4\vartheta_{12}) \sin(4\vartheta_{23}) \cos\delta_l [1 + s_{13}^2] \sin^2(\varphi_{21}/2)$$

$$- c_{13} \sin(2\vartheta_{12}) \sin(2\vartheta_{13}) \sin(2\vartheta_{23}) \sin\delta_l \sin(\varphi_{21})$$

$$- 2s_{13}^2 \sin^2(2\vartheta_{12}) \sin^2(2\vartheta_{23}) \cos(2\delta_l) \sin^2(\varphi_{21}/2) \}.$$

*Translated by R. Tyapaev*